\documentclass[aps,prl,reprint,twocolumn]{revtex4-1}

\usepackage{amssymb}
\usepackage{epsfig}
\usepackage{graphicx}
\usepackage{amsmath}
\usepackage{subfigure}
\usepackage{array,color}
\usepackage{dcolumn}
\usepackage{bm}

\begin{document}
\title{Topological Supersolidity of Dipolar Fermi Gases in a Spin-Dependent Optical Lattice}
\author{Huan-Yu Wang$^{1,2}$}

\author{Zhen Zheng$^3$}
\author{Lin Zhuang$^4$}

\author{Wu-Ming Liu$^{1,2}$ }
\email{wliu@iphy.ac.cn}
\affiliation{$^1$ Beijing National Laboratory for Condensed Matter Physics, Institute of Physics, Chinese Academy of Sciences, Beijing 100190, China}

\affiliation{$^2$ School of Physical Sciences, University of Chinese Academy of Sciences, Beijing 100190, China}

\affiliation{$^3$ Key Laboratory of Quantum Information, and Synergetic Innovation Center of Quantum Information and Quantum Physics, University of Science and Technology of China, Hefei, Anhui 230026, People's Republic of China}

\affiliation{$^4$ School of Physics, Sun Yat-Sen University, Guangzhou 510275, People’s Republic of China}
\begin{abstract}
We investigate topological supersolidity of  dipolar Fermi gases in a spin-dependent 2D optical lattice.
Numerical results show that the topological supersolid states can be synthesized via the combination of topological superfluid states  with the stripe order, where the topological superfluid states generated with dipolar interaction possess the $\Delta_{x}+i\Delta_{y}$ order, and it  is of D class topological classification. By adjusting the ratio  between hopping amplitude $t_{x}/t_{y}$ and  interaction strength $U$ with dipole orientation $\phi\approx \frac{\pi}{4}$, the system  will undergo  phase transitions among  the $p_{x}+ip_{y}$-wave topological superfluid state,  the $p$-wave superfluid state,  and the topological supersolid state.
The topological supersolid state  is proved to be stable by the positive sign of the inverse compressibility. We design an experimental protocol to realize the staggered next-next-nearest-neighbor hopping via the laser assisted tunneling technique, which is the key to synthesize topological supersolid states.
\end{abstract}
\maketitle

Supersolid (SS)  states can be defined as a combination of superfluid states with off-diagonal long range order and solid states with diagonal long range order, which was first emerged in the context of solid $\rm ^4He$ \cite{A.F. Andreev, C.V. Chester, A.J. Leggett, M.J. Bijlsma, Nikolay,L. Pollet, E. Kim}.  As a novel state of matter, supersolid states exhibit unconventional  properties and its realization methods  have long been an intriguing task  for theoretical and experimental physicists \cite{G. G. Batrouni, Pinaki Sengupta, Stefan Wessel, Erhai Zhao, Thierry Lahaye, Rittner ASC,  Ippei Danshita,O. Tieleman, X. Li, Jinwu Ye, L. He, AB, S. G. Bhongale, N. Henkel, N. Y. Yao, TF, Abraham,T. Mishra, Tian-Sheng Zeng,  Kwai-Kong Ng, W. Han, Raina J. Olsen, Wanzhou Zhang, G. R. M. Robb, Zhen-Kai Lu, Zhigang Wu, Renate Landig,  Xue-Feng Zhang, A. Keles, A Camacho Guardian, Geissler Andreas, Ji Guo Wang}. Recent works by Li \cite{Jun-Ru Li} and L$\rm\acute{e}$onard \cite{Leonard} realize periodical density modulation (stripes) in  Bose-Einstein condensates (BEC), which shed light on how to produce supersolid states in bosonic system. For fermionic system, the most promising candidate is dipolar Fermi gas due to its anisotropic features \cite{K. Aikawa, S. T. Chui}. By tilting the orientation of dipoles, dipolar interaction can be decoupled to a repulsive part and an attractive part in separated directions, which is the key to form the stripe and the superfluid order.  The two order parameters compete with each other, and supersolid states will arise at moderate interaction strength. However, the corresponding supersolid state is trivial in topology.

\begin{figure}[tbp]
\centering
\includegraphics[width=0.48\textwidth]{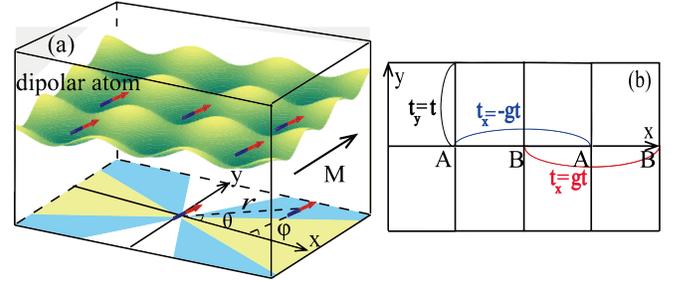}
\caption{(a) Dipolar Fermi gases in a 2D optical lattice with staggered next-next-nearest-neighbor hopping and its interaction projection. The dipoles form an angle $\phi$ with respect to $x$ axis, and $\theta$ is the angle between the dipole separation $\bf{r}$ and $x$ axis. $\rm\mathbf{M}$ is the external magnetic field. In the yellow area, the dipolar interaction decouples to an attractive part in $x$ direction and a repulsive part in $y$ direction.  In the blue area, the dipolar interaction decouples to attractive parts in both $x$ and $y$  directions. In the white area, the dipolar interaction decouples to an attractive part in $y$ direction and a repulsive part in $x$ direction. (b)  The schematic diagram of the anisotropic staggered next-next-nearest-neighbor hopping in the $x-y$ plane. A, B labels the sites of two sub-lattice.}
\label{fig1}
\end{figure}
%

To synthesize the topological supersolid state, we consider to combine topological superfluid
 states with stripe order, where the
topological superfluid state generated with spinless dipolar fermions possesses the $\Delta_{x}+i\Delta_{y}$ superfluid order \cite{Bo Liu} and it is of D class topological classification,  of which  the chiral symmetry is broken, and   the time reversal symmetry is broken by the relative phase $ \frac{\pi}{2}$ in the superfluid order. However, the conventional stripe order generated by the repulsive part of dipolar interaction can not coexist with the topological superfluid order.

In this Letter, we investigate the topological supersolid state of dipolar Fermi gases in a 2D optical lattice with staggered next-next-nearest-neighbor hopping, which is equivalent to a spin-dependent potential. We exhibit that our hopping formalism can cooperate with dipolar coupling between different optical sites to generate a new type of stripe order, which no longer relies on the repulsive part of dipolar interaction. When the dipolar interaction is decoupled to attractive parts in both $x$ and $y$ directions and the two parts are balanced in strength, the $p_{x}+ip_{y}$-wave topological superfluid state arises. Once the topological superfluid state is combined with the stripe order,  the topological supersolid state can be yielded, of which the topological features are characterized by the edge states. Furthermore, we prove that the topological supersolid state is stable via the positive sign of the inverse compressibility. The simplicity of our lattice model makes it reliable and feasible
in realization via  the laser assisted tunneling technique.

We consider $\rm^{161} Dy$ atoms confined in a 2D optical lattice, with the lattice potential given by $ V_{opt}(\mathrm{\mathbf{r}})=V_{0}[\sin^2(\pi x/a)+ \sin^2(\pi y/a )]$.  $V_{0}$ is the trap strength, and $a=450$ nm is the lattice constant \cite{Mingwu Lu}.
The next-next-nearest-neighbor hopping can be induced by the laser assisted tunneling technique \cite{M. Aidelsburger, Hirokazu Miyake} with the phase $\pi$ and $0$  staggeredly loaded along  the $x$ direction, which naturally leads to two effective types of atoms (A, B).  While along the $y$ direction, hopping is uniform, and   A, B types of atoms share the same lattice structure (see Fig. 1(b)). Thus, the lattice potential is equivalent to a spin-dependent one with A (B) regarded as pseudospin index respectively.  We assume $\rm^{161}Dy $  atoms are prepared in its lowest hyperfine nuclear spin states and spin degrees of freedom are frozen out. The system can be described by a spinless Fermi-Hubbard model,
\begin{equation}\label{eq:hamilton}
\begin{aligned}
\mathcal{H} = &\sum_{i}(-t_{x}c_{i}^{\dagger}c_{i+2\hat{e}_x}e^{i\pi i_{x}}-t_{y}c_{i}^{\dagger}c_{i+\hat{e}_{y}}\\
&-\mu c_{i}^{\dagger}c_{i}+\mathrm{H.c.})+\frac{1}{2}\sum_{i\neq j }V_{ij }c_{i}^{\dagger}c_{j}^{\dagger}c_{j}c_{i},
\end{aligned}
\end{equation}
where $ c_{i}^{\dagger}$ ($c_{i}$) creates (annihilates) a fermion at site $\mathrm{\mathbf{R}}_{i} $. $t_x$, $t_y$ is the hopping amplitude in $x$ and $y$ direction, and is denoted by $t_{x}\equiv t$, $t_{y}\equiv t$. Dipoles are aligned in parallel by external  magnetic field $\mathbf{M}$ into a direction which is inside the plane, and keeps an angle $\phi$ with respect to $x$ axis.
The interaction between dipole moments $\mathrm{\mathbf{d}}$  separated by $\mathrm{\mathbf{r}}$ is given by $V(\mathrm{\mathbf{r}})=\frac{d^{2}}{|r|^{3}}(1-3\cos(\phi-\theta)^2)$. A schematic picture of the system is shown in Fig. 1(a). Experimentally, our Hamiltonian provides three tunable parameters: (i) The hopping ratio $g$ can be controlled by adjusting the Rabi frequency in the laser-assisted tunneling technique.
(ii) The dimensionless coupling strength $U\equiv |d|^{2}/(ta^3)$, which characterizes $V_{ij}$,
can be changed by manipulating the strength of lattice trap. (iii) The orientation of dipoles $\phi$.
The controllable parameter set $\{g, U, \phi\}$ can give rise to a phase diagram with rich physics.

\begin{figure}[tbp],
\centering
\includegraphics[width=0.4\textwidth, height=0.3\textheight]{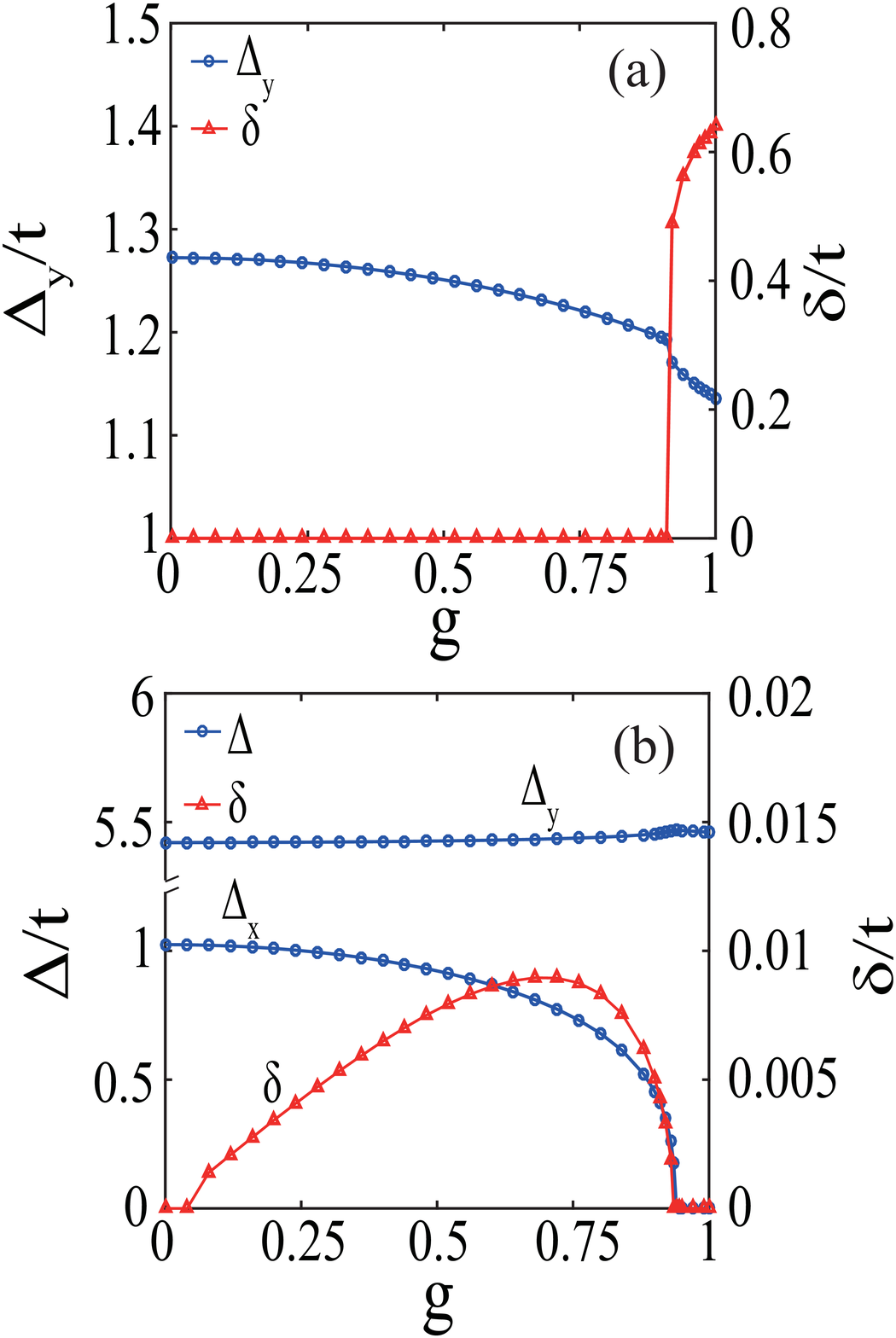}
\caption{The pseudogap as a function  of hopping ratio $g$ for different orientation $\phi$ of dipoles and interaction strength $U$. (a) Pairing gap $\Delta_{y}$ and stripe gap $\delta$ at $\phi=0.3\pi$, $U=5$. (b) Pairing gap $\Delta_{x}+i\Delta_{y}$ and stripe gap $\delta$ at $\phi=0.26\pi$, $U=30$.}\label{fig24}
\end{figure}

Our results are obtained by the Bogoliubov-de Gennes (BdG) approach. Within the mean field approximation \cite{Cheng Zhao}, the interaction  can be decoupled to $V_{HF}\!=\!\sum_{i\neq j}\Delta_{ij}c_{i}^{\dagger}c_{j}^{\dagger}+\Delta_{ij}^{\ast}c_{j}c_{i}+V_{ij}(\langle n_{i}\rangle c_{j}^{\dagger}c_{j}+\langle n_{j}\rangle c_{i}^{\dagger}c_{i})+E_{0}$,  where $E_{0}\!=\!\sum_{i\neq j} -\frac{\Delta_{ij}^2}{V_{ij}}-V_{ij}\langle n_{i} \rangle  \langle n_{j} \rangle   $.  We describe the superfluid order parameter as $\Delta_{ij}=V_{ij}\langle c_{i}c_{j} \rangle \delta_{j, i+\hat{e}_\lambda}$  with $\lambda=x, y$. To manifest the stripe order, we assume $\langle n_{i}\rangle= \frac{1}{2}+e^{i\bm{Q}\cdot \mathrm{\mathbf{R}}_{i}}C$, $|C|\leq\frac{1}{2}$. $ \bm{Q}$  represents the periodicity of density pattern. For the in plane dipoles, $\bm{Q}$  is assumed to be $\rm[\pi,0]$ \cite{Anne,K. Mikelson}. The free energy can be given by
$\Omega=TS$, and  the action S is of the form $S=\int d\tau \sum_{i}c_{i}^{\dagger}(\tau)\frac{\partial}{\partial\tau}c_{i}(\tau)+H$. By Fourier transformation and summing over the Matsubara frequency, we obtain $\Omega= E_0-T\sum_{n,\mathrm{\mathbf{k}}}l_n[1+exp(-E_{n}(\mathrm{\mathbf{k}})/T)]+\sum_{\mathrm{\mathbf{k}}}\widetilde{\kappa}$, where $E_{n}(\mathrm{\mathbf{k}})$ is the eigenvalues of $H_{BdG}(\mathrm{\mathbf{k}})$, and $\widetilde{\kappa}\!=\!-t_{y}\cos(k_{y}a)$. On the basis of $\Psi\!=\!(c_{\mathrm{\mathbf{k}}}, c_{\mathrm{\mathbf{k}}+\bm{Q}} , c_{\mathrm{\mathbf{-k}}}^{\dagger} , c_{\mathrm{\mathbf{-k-q}}}^{\dagger})$, we have $H_{BdG}(\mathrm{\mathbf{k}})$ can be described as:

%
\begin{equation}
H_\mathrm{BdG}(\mathrm{\mathbf{k}}) = \left(\begin{array}{cc}
A_{\mathrm{\mathbf{k}}}& D_{\mathrm{\mathbf{k}}} \\
D_{\mathrm{\mathbf{k}}}^\dag & -A_{\mathrm{\mathbf{k}}}
\end{array}\right)
\end{equation}
in which $A_{\mathrm{\mathbf{k}}}\!=\!-\frac{1}{2}t_y\cos(k_y a)I+\frac{1}{4}\big[\delta - 2t_x\cos(2k_xa)\big]\sigma_x$, and $D_{\mathrm{\mathbf{k}}}\!=\!\frac{-i}{2}\sum_{\lambda=x,y} \Delta_\lambda\sin(k_\lambda a)I$.
 Here, $\delta\!=\!V_{\bm{Q}}C$ and $V_{\bm{Q}}=\int \mathrm{d}\bm{r}\,V_{ij}(\mathrm{\mathbf{r}})e^{\mathrm{i}\bm{Q}\cdot \mathrm{\mathbf{r}}}$, which represents the stripe order. The self consistent gap equation  can be deduced via $\frac{\partial \Omega}{\partial \Delta}=0$, $\frac{\partial \Omega}{\partial \delta}=0$.  Need to mention that since one dipole interacts with many others, given the lattice scale large enough,  the effective dimension of the system can be so high that the fluctuation impacts are suppressed. Hence, we consider $500\times500$ lattice sites in calculating the Hartree term to ensure the validity of the mean field method.

The orientation of dipoles $\phi$, hopping ratio $g$ and interaction strength $U$ determine much of the physics of the system. When $\phi>0.3\pi$,  the dipolar interaction can be projected to an attractive part in $y$ direction, which results in a superfluid order, and a repulsive part in $x$ direction, which yields a stripe order. The two order parameters compete with each other and the system will undergo  phase transitions with the change of interacting strength $U$. For example, at $\phi\!=\!\frac{\pi}{2}$, superfluid order $\Delta_{y}$ is observed with modest interaction  strength $U$.  As $U$ is enhanced, stripe order gradually appears, and will dominate at large interaction strength.  The coexistence of the stripe order and the superfluid order defines a supersolid state.

Since we are only interested in supersolid states in the attractive interaction region, we restrict the dipoles to be arranged in the region $0.3\pi>\phi>0.2\pi$ (blue area in Fig. 1(a)).  When $\phi$ is about $0.3\pi$, the attractive interaction strength in $x$ direction is much weaker than that in $y$ direction, $V_{\lambda}=V_{ij}\delta_{i,{j+\hat{e}_{\lambda}}}$,  $\frac{V_{x}}{V_{y}}=0.0379$, as a consequence of which pairing order parameter can only be formed along $y$ direction.   By numerically locating the minimum of the free energy, we obtain the magnitude of  pseudogap as a function of hopping ratio $g$ at $ U=5$ (see Fig. 2(a)). It can be seen that the pairing gap $\Delta_{y}$   is insensitive to the hopping ratio g and  the stripe gap $\delta$ emerges for $g>0.92 $, indicating a  supersolid state. To investigate the origination of the stripe order with respect to g, we consider the expression of $A_{\mathrm{\mathbf{k}}}$ in Eq. (2), where term $\big[\delta - 2t_x\cos(2k_xa)\big]\sigma_x$ is equivalent to an effective $\mathrm{\mathbf{k}}$-dependent Zeeman field $h_{x}$ in $x$ direction. The dipolar interaction provides the coupling between different sites of the sub-lattice (or pseudospins) A, B, which is equivalent to a spin exchange interaction.  As a response to the magnetic field  $h_{x}$, pseudo spin wave will be formed and it corresponds to the stripe order. It is need to mention that when the dipolar interaction is absent, the system is in fact constructed by the two isolated sublattices A and B  without inter-sublattice coupling. For the sake of spin balance, the two sublattices A and B, although their intra-sublattice tunneling host a relative $\pi$ phase, up and down pseudospin will form a spacial uniform pattern. Thus the stripe order is formed spontaneously.
\begin{figure}[tbp],
\centering
\includegraphics[width=0.48\textwidth]{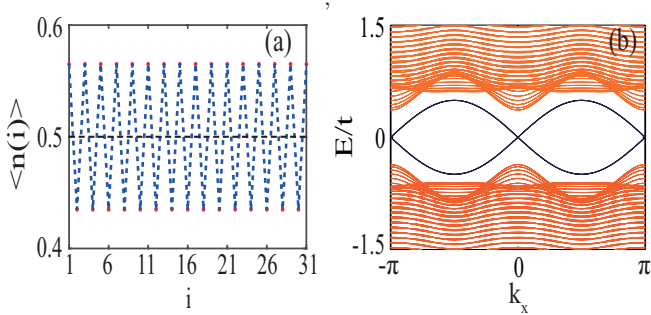}
\caption{ (a)  Density modulation of the topological supersolid states in real space: average particle density  at $i$th site along the $x$ direction with $g=0.3$, $\phi=0.26\pi$, $U=30$. (b) Edge states of topological supersolid states with  $g=0.3$, $\phi=0.26\pi$, $U=30$. Open boundary conditions are used in $y$ direction.}\label{fig24}
\end{figure}

As the orientation of dipoles $\phi$ reduces to $0.26\pi$, we have  $|\frac{V_{\bm{x}}}{V_{\bm{y}}}|=0.6829$, $|\frac{V_{\bm{Q}}}{V_{\bm{y}}}|=0.1989$. The superfluid order $\Delta_{x}$  appears, and is accompanied with  $\frac{\pi}{2}$ phase relative to that in $y$ direction. The superfluid  order $\Delta_{x}$, $\Delta_{y}$ and stripe order $\delta$ compete with each other as a function of hopping ratio $g$ and interaction strength $U$ (see Fig. 2(b)). Numerical results show that the stripe order can still be detected at large interaction strength and the simultaneous coexistence of the topological superfluid order $\Delta_{x}+i\Delta_{y}$ and the stripe order $\delta$  characterises a topological supersolid (TSS) state. Furthermore, it can be seen that with the increasing of the hopping ratio $g$, superfluid order $\Delta_{x}$ gradually shrinks due to the reduction of relative dimensionless interaction strength $ U/g$ along the $x$ direction and vanishes at $g=0.96$. Likewise, the stripe order reaches a peak at $g=0.7$, and will decay for $g>0.7$. To manifest the stripe order in topological supersolid states, we calculate average density at  $i$th site in the $x$ direction with $g=0.3$, $\phi=0.26\pi$, $U=30$ (see Fig. 3(a)). A periodical density modulation $n\langle i+1\rangle-n\langle i\rangle\!=\!C(e^{i\bm{Q}  \cdot \mathrm{\mathbf{R_{i+1}}}}-e^{i\bm{Q}\cdot \mathrm{\mathbf{R_{i}}}})=0.1302$ is observed in $x$ direction.

We map out the phase diagram as a function of  interaction strength U and hopping ratio $g$ at zero temperature as shown in Fig. 4. For $\phi=0.3\pi$ (see Fig. 4(a)), at modest interaction strength ($U<3.1$), stripe order is absent for all cases of $g$, and the system is a $p$-wave superfluid state. With the increasing of $U$, stripe order gradually emerges, and the system becomes a supersolid state. Phase diagram for $\phi\!=\!0.26\pi$ is shown in Fig. 4(b). Topological superfluid states and topological supersolid states exist at large coupling strength, and can transfer to each other with the tuning of hopping ratio $g$.
To investigate the topological classification of topological supersolid states, we denote the particle-hole reversal ($\rm\bm{\Xi}$) operator as: $\rm\bm{\Xi}\!=\!\sigma_{x}\bm{K}$, where $\rm\bm{K}$ is the complex conjugate operator. Thus, we have $ \rm\bm{\Xi} H_{BdG}(\mathrm{\mathbf{k}}) \bm{\Xi}^{-1}\!=\!-H_{BdG}(\mathrm{\mathbf{-k}})$.
Since $\rm\bm{\Xi}^2=1$ and the time-reversal symmetry is broken due to  the $\pi/2$ relative phase in the superfluid  order parameter, we clarify the topological supersolid state as D class, which can be characterized by the $\mathbb{Z}$ index \cite{A. P. Schnyder, M. Z. Hasan, X. L. Qi, C.-K. Chiu, Niu}. Edge states of topological supersolid states can be observed in Fig. 3(b).

\begin{figure}[tbp]
\centering
\includegraphics[width=0.48\textwidth]{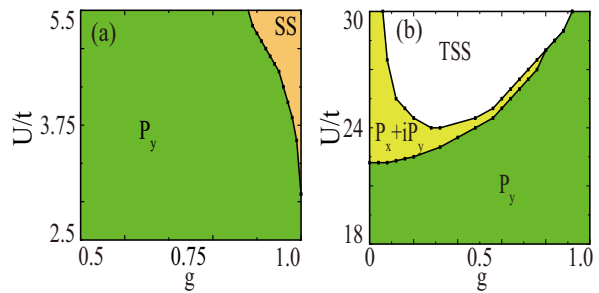}
\caption{The phase diagram as a function of hopping ratio $g$ and interaction strength $U$. (a)  The supersolid state (SS) and the p-wave superfluid state ($\rm P_{y}$)  survive in the attractive interaction region for $\phi\!=\!0.3\pi$. (b) The $p_{x}+ip_{y}$-wave topological superfluid state and the topological supersolid state (TSS) emerge at large interaction strength for $\phi=0.26\pi$. } \label{fig3}
\end{figure}

The stability of supersolid states in the attractive interaction region can be justified by the sign of inverse compressibility \cite{G.M. Bruun, Pedro M. Duarte, Shao-Jian Jiang}
\begin{equation}\label{eq:hamilton}
\begin{aligned}
\kappa^{-1}=\frac{\partial\mu}{\partial n}=-\frac{\partial^2\Omega}{\partial n^2}|_{n=n_{0}}, n_{0}=\frac{1}{2}.
\end{aligned}
\end{equation}
Numerical results for the compressibility as a function of hopping ratio $g$ with various interaction strength $U$ for $\phi\!=\!0.3\pi$ , $\phi=0.26\pi$ are shown in  Figs. 5(a), (b). It can be seen that the compressibility is always positive, which implies that supersolid states and topological supersolid states are stable. Besides, from Fig. 5(a), it can be seen that the compressibility increases with the augment of $g$ at first, and will encounter a reduction as  the stripe order arises. Indeed,  the effect of stripe order is equivalent to a  renormalized chemical potential $\widetilde{\mu}^{A (B)}_{i}$, which will retard the change of chemical potential. For $\phi\!=\!0.3\pi$, the stripe order dominates at large interaction strength. The energy band is fully gapped leading to a Mott insulator state, which is characterized by the vanished compressibility.

\begin{figure}[tbp]
\centering
\includegraphics[width=0.48\textwidth]{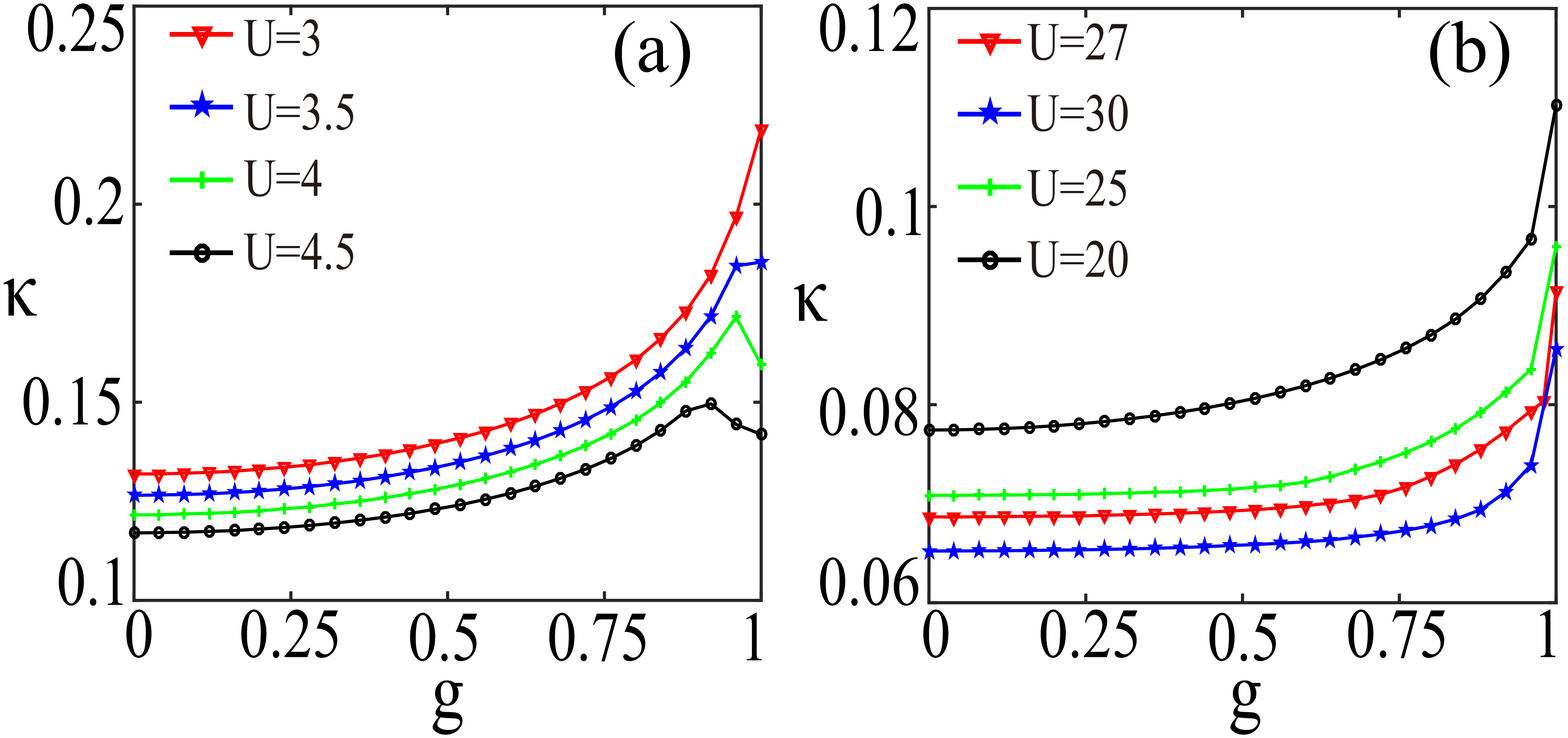}
\caption{ The compressibility $\kappa$  varies as a function of the hopping ratio $g$  with different interaction strength $U$ and dipole orientation $\phi$ . (a) $\phi=0.3\pi$ and $\kappa$ vanishes for $U=15$.  (b) $\phi=0.26\pi$.  } \label{fig52}
\end{figure}

%


\emph{Experiment realization}.--- In order to generate strong enough dipolar interaction strength, we consider the ground state $|F,mF\rangle=|21/2,-21/2\rangle$ of $\rm^{161}Dy$ dipolar atoms. The atoms are confined in a 2D optical lattice with lattice constant $a=450$ nm \cite{Mingwu Lu}, where the bare tunneling  is inhibited by the energy offset $\Delta$  and inter-sublattice tunneling is introduced via laser-assisted tunneling technique \cite{M. Aidelsburger, Hirokazu Miyake} (see Fig. 6(a)). Two far-detuned Raman beams ($k_{1},\omega_{1}$), ($k_{2},\omega_{2}$) are utilized to couple the next-next-nearest-neighbor sites with the difference of  laser frequency prefixed to be $\delta \omega=\omega_{1}-\omega_{2}\!=\!2\Delta$. The phase obtained along $x$ direction during the tunneling  process is $\rm\delta \mathrm{\mathbf{k}}\cdot \mathrm{\mathbf{R}}\!=\!m\theta_{x}+n\theta_{y}$, where $\theta_{x}\!=\! 2\pi|\sin(\varphi/2)|\cos(\gamma)  $, $\theta_{y}\!= \!2\pi|\sin(\varphi/2)|\sin(\gamma)$. Here $\varphi$ is the angle between two laser beems and  $\gamma$ is the angle between $\delta \mathrm{\mathbf{k}}$ and $x$ axis (see Fig. 6(b)).  To achieve our staggered next-next-nearest-neighbor hopping, we presume $\varphi\!=\!\pi/3$, and $\gamma\!=\!\pi$.
The hopping amplitude in $x$ direction is assumed to be $ t_{x}=\frac{\Omega}{2}\int dx w^{\ast}(x)e^{-2k_{x}x}w(x-2a)\int dy w^{\ast}(y)e^{-ik_{y}y}w(y)$, in which $w(x)$ is the wannier function. As a consequence, we can adjust the Rabi frequency to manipulate the ratio $g$ between the hopping amplitude $t_{x}, t_{y}$.  The supersolid  state can be detected through  time of flight measurements \cite{M. Greiner, S. Folling, T. Rom, I. B. Spielman, Zhigang Wu}.

In conclusion, we investigate topological supersolid states of dipolar Fermi gases in a 2D optical lattice with staggered next-next-nearest-neighbor hopping. The novel state obtained here, topological supersolid  state, paves a way to realize topological perfect crystal, and it is also of significant potential for quantum computing. We also utilize the inverse compressibility to prove that topological supersolid states will not collapse at large coupling strength. We propose the experiment setup to simulate the topological supersolid state and its corresponding topological phase transition.

\begin{figure}
\includegraphics[width= 0.48\textwidth]{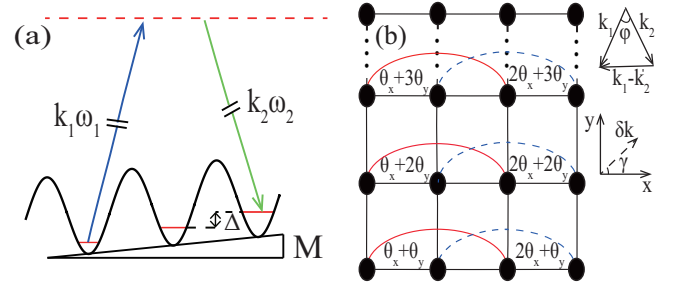}
\caption{(a) Experimental setup for staggered next-next-nearest-neighbor tunneling via Raman laser assisted tunneling. $(k_{1},\omega_{1})$, $(k_{2},\omega_{2})$ are two far detuned Raman lasers, $\Delta$ is the energy offset induced by gradient magnetic field $\rm\bm{M}$. (b) The phase accumulated in the Raman laser assisted tunneling process. $ \varphi $ is  the angle between two laser beams, $\gamma$ is the angle between $\rm\delta \mathrm{\mathbf{k}}$ and $x$ axis.} \label{fig5}
\end{figure}

We are grateful to Lan Yin, Xu-Bo Zou,  and Ru-Quan Wang for helpful discussions. This work was supported by the NKRDP under grants Nos. 2016YFA0301500,  NSFC under grants Nos. 11434015, 61227902, 61378017, SKLQOQOD under grants No. KF201403, SPRPCAS under grants No. XDB01020300, XDB21030300. Z.Z. acknowledges support from National Natural Science Foundation of China Grants No. 11474271 and No. 11674305, National Postdoctoral Program for Innovative Talents of China Grant No. BX201600147.

%

\end{document}